\documentclass[twocolumn]{revtex4}

\usepackage{epsfig}

\newcommand{\FF}{\mathcal{F}}

\newcommand{\Sol}{\mathcal{S}}

\newcommand{\bF}{{\mathbf{F}}}

\begin{document}

\title{Force indeterminacy in the jammed state of hard disks}

\date{\today}

\author{Tam\'as Unger}
\affiliation{Dept.\ of Theoretical Physics, Budapest University of
  Technology and Economics, H-1111 Budapest, Hungary}
\affiliation{Institute of Physics, University Duisburg-Essen, D-47048,
  Duisburg, Germany} 
\author{J\'anos Kert\'esz}
\affiliation{Dept.\ of Theoretical Physics, Budapest University of
  Technology and Economics, H-1111 Budapest, Hungary}
\author{Dietrich E. Wolf}
\affiliation{Institute of Physics, University Duisburg-Essen, D-47048,
  Duisburg, Germany} 

\begin{abstract} 
  Granular packings of hard discs are investigated by means of contact
  dynamics which is an appropriate technique to explore the allowed
  force-realizations in the space of contact forces. Configurations
  are generated for given values of the friction coefficient, and then
  an ensemble of equilibrium forces is found for fixed contacts. We
  study the force fluctuations within this ensemble.  In the limit of
  zero friction the fluctuations vanish in accordance with the
  isostaticity of the packing.  The magnitude of the fluctuations has
  a non-monotonous friction dependence.  The increase for small
  friction can be attributed to the opening of the angle of the
  Coulomb cone, while the decrease as friction increases is due to the
  reduction of connectivity of the contact-network, leading to local,
  independent clusters of indeterminacy. We discuss the relevance of
  indeterminacy to packings of deformable particles and to the
  mechanical response properties.
\end{abstract}

\maketitle

Jamming \cite{Makse04} has been in the focus of recent studies because it
occurs in a great variety of phenomena like structural and spin glasses,
colloidal systems, vehicular traffic and granular media.  The
characterization of the jammed state is therefore crucial and can perhaps
be best achieved in granular systems.  Many intriguing properties of
granular packings originate from the microscopic force transmission through
a contact structure, where non-linearity and disorder are known to be
crucial. It is an essential but not resolved question how the highly
inhomogeneous force-network influences the macroscopic stress transmission
in dense granular media.

Since the deformations of the grains are usually much smaller than
their
size, a very useful reference system for granular
matter is that of rigid (undeformable) particles
\cite{Edwards89,Mehta89,Makse04,Radjai96}.
% A further
%possible simplification is 
%the assumption of spherical (3d) or disk-shaped (2d) particles. 
%It is known that random packings of frictionless rigid disks or spheres
%exhibit an \emph{isostatic} force-carrying structure
%\cite{Moukarzel98,JNRoux00,Tkachenko99}, where all contact forces are fully
%determined by the equilibrium conditions for each
%grain. It is less clear what happens in the presence of
%friction. Simulations with soft particles and an extrapolation to the hard
%particle limit show that the problem is \emph{hyperstatic}
It is known that random packings of frictional rigid disks or spheres
exhibit a \emph{hyperstatic} structure
\cite{Silbert02,Unger03b,JNRoux00}: the number of the linear equilibrium
equations of the grains, which relate the unknown contact forces to the
external load, is
too small to determine the contact forces uniquely.
%The isostaticity is often characterized by the average
%coordination number of the grains. Matching the number of the equations and
%unknowns provide  critical coordination number $z_c = 4$ ($3$) for
%frictionless (frictional) disks. Note, however, that the question of
%determinacy of the forces is more subtle and in many cases can not be
%judged just by counting contacts (for an extensive study see \cite{JNRoux00}.)
Therefore many mechanically admissible force-networks are possible in
the same packing geometry and for the same external load, which define an
\emph{ensemble} of force-configurations. 

This ensemble recently has
received much attention
\cite{Bouchaud03,Snoijer04,Unger03b,Ostojic04,Elperin98,Clement98,Moreau04,McNamara04}
%in the context of granular media.
due to the idea that
some macroscopic properties of jammed granular systems
% (macroscopic stress,
%response properties, force distribution) 
can be derived
based on an ensemble average over the admissible force-states
\cite{Bouchaud03}. The determination of force distribution in
\cite{Snoijer04} or Green function in \cite{Ostojic04} are based on this
approach.

% where the average is taken over force and also over
%packing configurations corresponding to one macroscopic state \cite{Edwards89,Mehta89,Makse04}. The
%important question addressed in this Letter is whether the
%force-state obtained by construction of a packing is special among other
%admissible states or the solutions are equivalent. Edwards' micro-canonical
%assumption suggests for the restricted case that the proper measure is
%uniform in the space of the contact forces, which is the starting point in
%\cite{Snoijer03,Ostojic04}.

Another interesting aspect of the force-ensemble is related
to the behavior of the system under
external perturbations. Packing structures where contact
forces are unique or strongly
restricted appear to be fragile:
slight change of the load can cause rearrangements of the particles
\cite{Cates98,Combe00}. The question arises whether a packing
that exhibits many possible realizations of equilibrium forces becomes more
robust against perturbations.
 
%related to the force transmission and to the response of the system to
%external perturbations: The freedom in choosing a realization of force
%equilibrium from a large set of possible solutions 
%suggests some robustness in the sense that a
%small change of the external forces can be resolved without changing the
%packing structure by just altering the contact forces. 
%On the other hand if the forces are strongly
%restricted one would expect that the structure is more fragile and a
%slight change of the load causes rearrangements of the particles
%\cite{Cates98,Combe00}.

The results of this Letter provide nontrivial information also for
packings of deformable particles: The actual network of contact forces
(which is uniquely determined by the elastic deformations) must be
contained in the force-ensemble calculated for the same contact
geometry assuming the particles (in their deformed shape) would be
perfectly rigid. Moreover, for a finite system of sufficiently rigid
particles the contact geometry can be arbitrarily close to the
ideal one obtained for perfect rigidity. Which of the solutions in the
force ensemble is realized, depends e.g.\ on the elasticity
%and the preparation history.
of the individual grains. Here we address the question, how strong the
restrictions provided by the force ensemble are.

Again another but closely related issue is that of hard particle
simulations, where the dynamics
is seemingly ambiguous due to the indeterminacy of forces \cite{Moreau04}.

The above problems indicate the significance of the force-ensemble,
however very little is known about its properties. In this Letter
some characteristics of the ensemble are revealed, where emphasis is put
on the influence of friction.

In the recent literature 
\cite{Snoijer04,Ostojic04} it was suggested that all 
elements in the ensemble of 
admissible force configurations are realized with equal probability.
This microcanonical approach
%(uniform) distribution, which 
can be regarded as a restricted version  \footnote{In 
  Edwards' thermodynamic theory the ensemble average is taken over force-
  and also over packing-configurations.} 
of the Edwards ensemble \cite{Edwards89,Mehta89,Makse04}.
In the following 
we also address the validity of this assumption.

Let us consider $n$ rigid, cohesionless disks. 
A configuration of the contact forces $\left\{ \bF_i \right\}$ (where $i$
is the 
contact index) is called admissible or a solution
% of the problem 
if two
conditions are fulfilled: the \emph{equilibrium} and the \emph{Coulomb}
conditions. The 
first one requires force and torque balance at each grain, while the Coulomb
condition reads:
%requires the following relation between the normal and tangential force at
%each contact:
\begin{equation}
  \label{Coulomb}
  \left| \left( \bF_i \right)_t \right| \le \mu \left( \bF_i \right)_n \ 
\end{equation}
for the normal and tangential force at each contact, 
where $\mu$ is the friction coefficient. For $\mu > 0 $ no additional condition
is needed to exclude tensile forces.

% ensures also that the contact forces are \emph{compressive}, but for
%$\mu=0$ the inequality 
%(\ref{Coulomb}) should be replaced by $0 \le \left( \bF_i \right)_n$  (in
%order to exclude tensile forces).

Next we show that the solutions form a convex set.  The space of
contact forces $\FF$ is defined (for fixed contact network) as an
$N_c\times d$ dimensional vector space, where each point represents a
force-configuration $\left\{ \bF_i \right\}$. $N_c$ is the number of
contacts, and $d$ the space dimension (i.e. each contact force
component represents one degree of freedom). Let $\Sol$ be the
\emph{subset of admissible states} in $\FF$ under some fixed external
forces. For a regular packing of disks $\Sol$ is known to be a convex
polyhedron \cite{Elperin98} but it is easy to see that \emph{convexity
  is satisfied in any case}: shape of the particles, disorder,
dimensionality or friction do not matter.  Convexity means that if
$\left\{ \bF_i \right\}$ and $\left\{ \bF_i + \Delta \bF_i \right\}$
are solutions then $\left\{ \bF_i + \lambda \Delta \bF_i \right\}$ is
a solution as well for $0 \le \lambda \le 1$. First, the equilibrium
condition holds: Both given force-configurations provide equilibrium
against the external load, thus their difference $\left\{ \Delta \bF_i
\right\}$ corresponds to zero load and exerts no total force or torque
on the particles. Therefore it can be scaled freely (unrestricted
$\lambda$) and added to an admissible state, that does not violate
the linear equilibrium equations. Second, the Coulomb condition is satisfied
simply because for each contact $i$ the d-dimensional Coulomb ``cone'' is a
convex set and therefore must contain the component $\bF_i+\lambda
\Delta \bF_i$, with $0\leq \lambda \leq 1$.

%it defines
%a convex set: since this Coulomb ``cone'' contains the
%forces $\bF_i$ and $\bF_i+\Delta \bF_i$ the points in between
%$\bF_i+\lambda \Delta \bF_i$, $0 \le \lambda \le 1$ also fulfill the
%condition~(\ref{Coulomb}).
 
 The solution set $\Sol$ reflects basically the properties of the
 contact-network, therefore when studying $\Sol$ it is crucial what
 kind of packing structure is considered. In real processes which lead
 to jamming, the microscopic structure is not prescribed but develops
 spontaneously up to the point, where further rearrangements against
 outer driving forces are blocked. This \emph{self-organized} texture
 is an important feature of granular materials \cite{Cates98} which is
 disregarded in models using, e.g., regular arrangements
 \cite{Elperin98,Clement98}.  Therefore the packings studied below
 were constructed with discrete element simulations where the
 particles obeying Newton's dynamics build up the contact-network
 in a compression process. In these jammed configurations
 we search for various solutions of the contact forces and study the
 influence of friction on the properties of $\Sol$.

%\begin{figure}[t]
%\centerline{\epsfig{figure=demo50_gray.eps,width=0.8\linewidth}}
%\caption{Allocation of the subsystem demonstrated in a packing of $50$
%  disks. (a) The jammed state. (b) The subsystem with fixed boundary
%  forces. The drawn network represents one admissible force-configuration
%  indicating the strength of the normal forces with the line width.}
%\label{fig:demo}
%\end{figure}

A detailed description of our method of constructing the packings and
exploring admissible force-configurations can be found in \cite{Unger03b},
here only a short review is given. With the help of the contact dynamics
algorithm \cite{Jean99,Unger03a} a 2D system of $200$ rigid disks is
compressed along the 
%$y$-
vertical axis between two horizontal plates. Horizontally
periodic boundary conditions are applied, gravity is set to zero, disk radii
are uniformly distributed between $R$ and $2 R$, the horizontal system width
is $42 R$. We wait till the packing jams (relaxes into equilibrium)
under the constant force of compression. Then, 
%in order to get rid of
to avoid the effect of the straight plates, only the middle part of the
static configuration is considered for further investigation: this is a
horizontal slice of height $28 R$ throughout the whole width in the bulk
away from the plates. We retain the contact forces at the top and bottom
perimeter of the slice as fixed boundary forces, thus they provide the
external load on the system. The plates and the disks outside the slice can
be left away.

% and cut out of the bulk away from the plates: a horizontal
%slice of height $28 R$ is taken throughout the whole width and used for
%further investigation. The parts below and above the slice were neglected
%after that.

%Together with this bulk material we kept also the contact forces of the
%equilibrium state, including the contact forces between the slice and the
%neglected part of the system. These latter forces (acting now at the perimeter
%of the slice) were regarded as fix boundary forces and provided the external
%load on the system. FIXED POSITIONS,..

% in zero gravity.  One
%plate is fixed while the other one is pushed with an external force
%(Fig.~\ref{fig:demo}.a). In $x$-direction a periodic boundary
%condition is applied. The disk radii are randomly chosen and uniformly
%distributed between $R$ and $2 R$, the width of the system $L_x=42
%R$. After the packing relaxed into equilibrium we stop the simulation and
%fix the particle positions. For the further investigation a subsystem is
%chosen in the bulk away from the plates: It is defined by two straight
%lines  $28 R$ apart parallel to
%the plates (Fig.~\ref{fig:demo}.a) containing disks with centers
%between the lines. We fix the contact forces between inner and outer disks
%(the forces obtained by the compression process) and regard them as the
%external load on this piece of bulk material
%(Fig.~\ref{fig:demo}.b). The outer disks do not play any role in the
%calculations.  

After that the exploration of the admissible force-solutions follows for
this fixed arrangement of disks.  We start with the force state that
appeared at the jamming and perturb all contact forces randomly
\footnote{We add random tangential and normal components to the contact
forces, which are chosen uniformly from 
$[-\langle F_n\rangle,\langle F_n\rangle]$, where $\langle F_n\rangle$ is
the average normal force in the system.}, which leads out of
equilibrium and violates the Coulomb condition. This perturbed state serves
as the input for the Gauss-Seidel-like iterative solver of the contact
dynamics method. This iterative algorithm lets the forces relax into a
consistent state, providing a (possibly) new solution
\cite{Unger03a,Unger03b}. The perturbation and relaxation can be repeated
many times always starting from the last solution (a kind of random walk in
the force space); in that way it is possible to \emph{sample points} from
$\Sol$. 

Based on this collection of force solutions
% is representative for $\Sol$, 
we can assess the differences between admissible states and study the 
problem of force indeterminacy.
The main feature of $\Sol$ that we found in
these self-organized structures is that the admissible force-networks are
%Although in the presence of friction many
%different force networks are admissible for the same particle configuration
%and a given external load, we found that they are all 
rather similar: The pattern of strong force lines changes little from
one realization to the other, showing that the 
%structure of the 
contact-network \emph{imposes strong restrictions} on the force-configuration.

For 
each contact force $\bF_i$ its variance $(\delta F_i)^2$ is calculated over the
measured realizations. The ratio
\begin{equation}
  \label{eta}
%  \eta=  \frac{\langle |\delta \bF_i|\rangle} {\langle \bF \rangle} \ 
  \eta=  {\langle \delta F\rangle}/ {\langle |\bF| \rangle} \ 
\end{equation}
represents the ensemble fluctuation in $\Sol$, thus it can be regarded
as a \emph{measure of ambiguity of the forces}. $\langle \cdot
\rangle$ means the average over all contacts.  The \emph{force
  ambiguity} $\eta$ has to be distinguished from the \emph{degree of
  indeterminacy} which refers to the dimension of the affine subspace
of force configurations solving the equilibrium
conditions (without the restrictions due to the Coulomb cones).
% into account).

\begin{figure}[t]
\centerline{\epsfig{figure=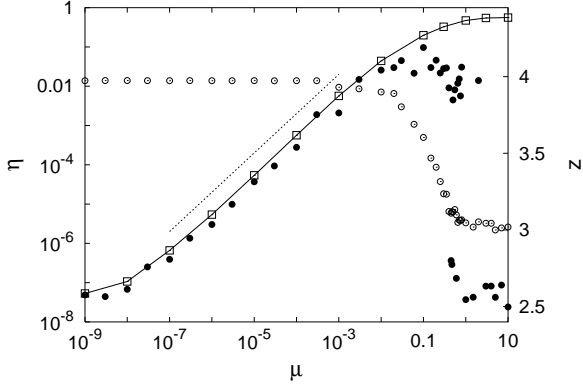,width=0.95\linewidth}}
\caption{Force ambiguity $\eta$ (full circles) and average coordination number
$z$ (open circles) as functions of the friction coefficient $\mu$.
For comparison,
squares connected by line show the $\eta$ values for a
configuration of disks that was constructed without friction.
%Ensemble fluctuations of contact forces measured for various
%coefficients of friction $\mu$. The packings were constructed either with the
%corresponding $\mu$ (dots) or without friction (squares connected by
%the solid line). Circles denote the
%average coordination number $z$.
}
\label{fig:fluct}
\end{figure}

%In order 
To investigate the effect of friction a
\emph{new} packing is constructed for each value of $\mu$ before sampling
the solutions.
The force ambiguity $\eta$ is plotted in Fig.~\ref{fig:fluct} (full circles). 
Values of
$\eta$ around $10^{-7}$ reflect the accuracy level of our calculation and the
corresponding force-configurations can be regarded as identical with this
tolerance. In the zero friction limit the force ambiguity disappears
\emph{confirming isostaticity} of frictionless packings
\cite{Moukarzel98,JNRoux00,Tkachenko99}.
For small $\mu$  the force ambiguity
grows proportionally with friction, however for larger $\mu$ it decreases 
again. The largest ambiguity of the
forces is found around $\mu \approx 0.1$.  Despite the further opening of the
Coulomb angle fluctuations are getting smaller, even fully determined
states are found for strong friction.
%which occurs in an
%interesting way: Around $\mu= 1$ it is a matter of chance whether the
%contact texture exhibits force fluctuations or not, with more
%inclination towards determined states (where
%only one admissible force state exists) as $\mu$ is increased.

The behavior of $\eta$ results from \emph{two competing effects}: first,
increasing 
friction provides larger freedom locally for the tangential forces,
%(Eq.~\ref{Coulomb}) 
second, it also stabilizes the system in a less dense
state \cite{Kadau03} causing lower connectivity of the 
contact-network (open circles in
Fig.~\ref{fig:fluct}), which reduces force ambiguity. One can separate the
two effects by 
fixing the configuration and letting the Coulomb angle alone
influence $\eta$: We generated one packing
without friction but switched on friction before sampling
force-configurations. The results obtained this way (squares in
Fig.~\ref{fig:fluct}) provide monotonously increasing
fluctuations, as expected. Compared to the original data (full circles)
% a good agreement is
%found on the left side of the figure and large 
deviations appear only on
the right side of the figure, where the changes in the connectivity become
important, while the behavior on the left side 
% which indicates that the small friction limit 
is governed by the first effect. For small $\mu$ 
%, and  only
%for large $\mu$. Accordingly one can assume that for a tiny friction
the average coordination number of the configuration is essentially the same 
as in the frictionless case, where from isostaticity $N_c \approx 2n$ follows.
This gives us the \emph{degree of indeterminacy}:
$2N_c - 3n \approx N_c/2 $, since there are two unknown force
components per contact and three equations per disk due to force and torque
balance. Thus we conclude that for tiny friction there is a \emph{small but
high-dimensional} set of force-solutions in the $2 N_c$ dimensional
force-space, and its size goes to zero with vanishing
friction. 
Similarly for spheres in three dimensions one obtains an 
$N_c$-dimensional solution set $\Sol$ within a $3N_c$-dimensional force space 
$\FF$.

%
%frictional case force and  torque balance provide three equations per disk 
%compared to two unknown force components per contact, so that
%the  in the $2N_c$ dimensional
%force-space. Therefore the set of the solutions is rather \emph{high
%dimensional} in the small friction limit. This does not contradict
%to the linearly vanishing fluctuations, which comes from the
%``diameter'' of $\Sol$ confined by Eq.~(\ref{Coulomb}).

For large $\mu$ the
dimension of $\Sol$
%%the solution set 
%caused by the
%low connectivity of the texture. 
%Here the Coulomb angle provides wide boundaries but the value of
%$d(\EE)$
is strongly reduced due to the decreasing number of contacts. In our
small system we found that
% the dimension of 
$\text{dim}(\Sol)$ can reach even zero,
allowing only one \emph{single force-configuration}.
%  finally reaches zero with high
%probability above $\mu=1$, resulting in static determinacy of forces.
%These geometries are the zero-gravity variant
%\footnote{The absence of gravity permits particles being without
%contacts.} 
This case corresponds to the marginal rigidity state found in experiments
\cite{Blumenfeld01}. 

\begin{figure}[t]
\centerline{%
  \epsfig{figure=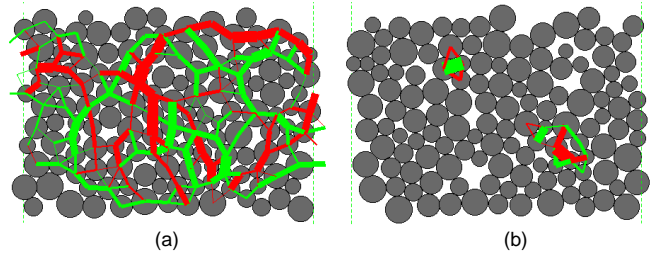,width=1\linewidth}}
\caption{(Color online) The difference between two admissible force-networks for (a)
$\mu=0.1$ (b) $\mu=0.5$. Only normal force differences are indicated with
  different colors depending on their sign.} 
\label{fig:localiz}
\end{figure}

The regression of the degrees of freedom occurs in an interesting way:
the \emph{indeterminacy gets localized} in space into small 
subgraphs of the contact-network, which are surrounded by
determined forces, i.e.\ a relatively large ambiguity is present but only in
a small part of the system (Fig.~\ref{fig:localiz}.b). The pattern of the
fluctuation-bearing contacts can be visualized 
%in Fig.~(\ref{fig:localiz}.b) 
%for a packing produced with $\mu=0.5$ where 
by plotting the difference between any two admissible
force-configurations. We found the same subgraphs as in
Fig.~\ref{fig:localiz}.b also for other arrangements of boundary forces,
showing that this indeterminacy-pattern is indeed a property of the
packing texture. Each of the two subgraphs shown in Fig.~\ref{fig:localiz}.b
is statically indeterminate, carries only one degree of freedom and cannot
be reduced further because the deletion of one particle or one contact
would cancel the internal indeterminacy. We call such subgraphs 
\emph{elementary clusters}. They can be regarded as geometric units of 
indeterminacy.
%
%the solution set $\Sol$ in the case shown in Fig.\ref{fig:localiz}.b is
%simply a rectangle meaning that each of the 
%two islands has one degree of freedom. We call such geometric units of
%fluctuations \emph{elementary clusters}. They are statically indeterminate
%but can not be reduced further (i.e.\ the deletion of one contact cancels
%the internal indeterminacy). In 2D the smallest elementary cluster for strictly
%convex particles is represented by the smaller graph in
%Fig.~(\ref{fig:localiz}.b) (consists of $4$ particles and $5$ contacts). 

If the connectivity is high the formation of elementary clusters is
more probable, which suggest the following picture: For small friction
many overlapping elementary clusters are formed so that two admissible
solutions generically differ throughout the system
(Fig.~\ref{fig:localiz}.a). As $N_c$ is reduced the density of the 
elementary clusters $\rho$ decreases and the indeterminacy gets localized
into small separated domains. Around $\mu=1$ the density $\rho$ becomes 
so small that
only a few elementary clusters are present due to the finite system size.   
This explains the strong scattering of the data for 
$\eta$ in Fig.~\ref{fig:fluct}.

The spatial localization raises the question of a percolation
transition. In case of small $\rho$ the separated domains carry
force-fluctuations \emph{independently} of 
each other, therefore we think that $\eta$ becomes a well defined intensive
quantity for large systems. However if the indeterminacy percolates through
the system the overlapping elementary clusters provide fluctuating boundary
forces for each other, thus the indeterminacy of forces is enhanced with
growing system size. Simulations up to $500$ particles show this \emph{size
dependence}, but it is not clear what happens in the thermodynamic limit.

\begin{figure}[t]
\centerline{%
  \epsfig{figure=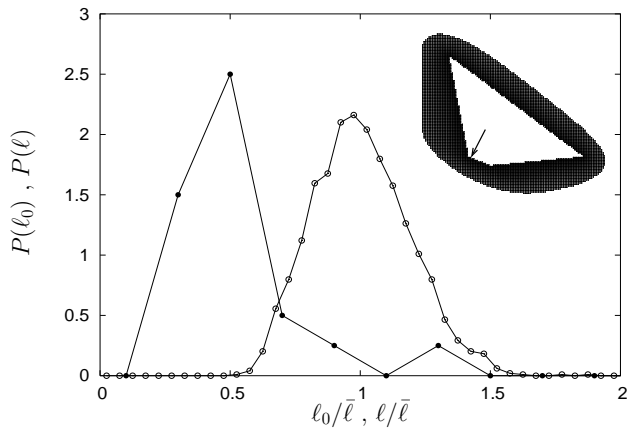,width=1\linewidth}}
\caption{Histograms of the distributions of normalized distances of 
dynamically generated (full circles) and randomly sampled (open circles) 
points in the sets $\Sol$ for $\mu=0.01$. The inset shows a two dimensional 
cross section of a high dimensional solution set. The dynamically constructed
force state is marked by the arrow. The white area belongs to $\Sol$, while
outside $\Sol$ the gray 
scale indicates the violation of the Coulomb condition (darker means smaller 
violation).} 
\label{fig:distance_histogram}
\end{figure}

Finally we investigate the dynamically created force-configuration
$\left\{ \bF_{i,0} \right\}$, which is determined by the construction
history. Our findings indicate that this state is \emph{more
  ``central''} than typical points in the solution set: We generate 20
initial configurations with $\mu=0.01$ and sample for each of them
$100$ points randomly in $\Sol$. Their (vectorial) average is regarded
as the center of $\Sol$. Then we measure the Euclidean distances
$\ell$ of the sampled points from the center. The histogram of the
distances in units of their average $\bar\ell$ is shown in
Fig.~\ref{fig:distance_histogram} together with the histogram of the
distances $\ell_0$ of the initial, dynamically generated 20 points from the
centers of the corresponding sets $\Sol$.  The two histograms clearly
indicate that the initial points are closer to the center on average
than the randomly sampled ones. Assuming that the distribution of the
random sampling of $\Sol$ is close to a uniform one,
%\footnote{This
%  assumption is supported by our results for a configuration with two
%  elementary clusters (see Fig.~\ref{fig:localiz}).}
 we conclude that the force configurations of the dynamically generated
jammed states are not uniformly distributed in the set $\Sol$.

That the original force configuration is ``closer to the center'' is
not in contradiction to the fact that we always find it at the edge of
two dimensional cross sections of the {\em high dimensional} solution
set $\Sol$ (see inset of Fig.~\ref{fig:distance_histogram}).  We
suggest the following physical picture: A contact with large
mobilization of friction ($F_t/\mu F_n \approx 1$) is less stable
against perturbations. Near the end of the relaxation process small
collisions ``shake'' the already established contacts reducing the
possibility that the contact remains on the verge of sliding.
However, the system comes to rest finally by the marginal fulfillment
of the Coulomb criterion at {\em some} contacts.

Our results show a significant difference between distributions of the
solutions sampled by the random walks plus relaxation and of those relaxed
physically. The uniformity of the (unbiased) random walk based sampling
cannot be proved due to the high dimensionality of the problem, however,
the distance distribution of the points should be rather robust just
because of this high dimensionality.
Therefore we consider the observed
discrepancy though not as a proof but as a strong indication of the
violation of the microcanonical assumption for the physically realized
solutions.

It is expected that the
ambiguity of forces for a given geometry has implications for the
mechanical behavior. 
We regard the following preliminary result as an indication of such an effect.
For a horizontal layer of hard disks settled under 
gravity we applied a point-force downwards on the free surface, just strong 
enough to cause local rearrangement. We measured the depth of the rearrangement
zone and obtained non-monotonous dependence on $\mu$: It is larger for small
and large friction coefficients, and has a minimum at $\mu \approx 0.1$, right 
where $\eta$ reaches its maximum.

%Acknowledgement 
This research was partially supported by DFG grant SFB 445, 
BMBF/OM grant HUN 02/011, OTKA T035028, and the Center
for Applied Mathematics and Computational Physics of the BUTE.

\bibliography{granu}

\end{document}